# Determining the magnitudes and spectral types of the components of the binary Mira X Ophiuchi

David Boyd
*Variable Star Section, British Astronomical Association, [davidboyd@orion.me.uk]*

**Abstract**

Several values for the magnitudes and spectral types of the components of the binary Mira X Ophiuchi have been published in the literature over the last century. Analysis of new photometry and spectroscopy of the star between 2016 May and 2016 December indicates that the V magnitude of the constant star is 9.0 and its spectral type K1III. The spectral type of the Mira changed from M6III at maximum to M7III as it faded and passed through minimum. The Mira's V magnitude varied between 6.47 at maximum and 9.83 at minimum, a range of 3.36 magnitudes.

**What do(n't) we know about X Oph?**

X Ophiuchi was first found to be variable by Espin observing at Darlington in 1886 (1). In 1900 Hussey discovered the star to be a visual double using a power of 1000 on the 36-in refractor at Lick Observatory (2) and measured the separation as 0.22 arcsec. In the Annals of the Harvard College Observatory in 1907 (3) Cannon reported its magnitude range as 6.5 to 9.0, the period of its brightness variation as 335 days and its spectral type as Md on the Draper spectral classification indicating the presence of Hγ and Hδ emission lines. Van Biesbroeck, observing at Yerkes Observatory in 1920-24 (4), deduced from brightness estimates that the more northerly component was the variable, a conclusion later independently confirmed by Gingrich at Mt Wilson Observatory (5). Van Biesbroeck visually estimated the magnitude of the southern constant star as 8.9 and the minimum magnitude reached by the variable component as 9.9, but in any case not fainter than 10.0.

In 1921 Merrill at Mt Wilson Observatory (6) reported that the Hγ and Hδ emission lines peaked around maximum light and disappeared as the star faded. Two years later (7) he reported the spectral type as M6e at maximum becoming K0 at minimum as the constant component dominated. Spectral type M6e is an extension of the Draper classification scheme adopted by the IAU in 1922 with the letter e indicating the presence of emission lines. Merrill gave the visual magnitude of the constant star as 8.9 and the visual range of the variable as 6.8 to 12. Pettit & Nicholson (8) found a visual magnitude at minimum of 11.5 for the variable based on radiometric measurements and a magnitude of 8.9 for its constant companion.

In a comprehensive review of current knowledge about X Oph, Fernie (9) reported an analysis of the relative intensities of spectral lines which gave the spectral type of the constant star as K1III. Based on several assumptions, he obtained a V magnitude for this star of 8.51 and a maximum V magnitude of the variable as 7.12. He also derived the distance to X Oph as 240±35 pc and its colour excess as E(B-V) = 0.15.

Based on the MK (Morgan and Keenan) spectral classification system using relative intensities of specific absorption and emission features, Keenan (10, 11) reported a range of spectral types for X Oph between M6e and M8e+K. These observations covered the full magnitude range of X Oph and showed the spectral type becoming later as the star faded with the M-type spectrum almost disappearing relative to the K-type spectrum at minimum light.

Skiff's Catalogue of Stellar Spectral Classifications available in Vizier (12) contains references to several papers reporting spectral types for X Oph ranging from M4e to M9. The current Simbad (13) entry for X Oph lists its spectral type as M0-8e+K2III, M0-8e being the modern equivalent of the old Md classification. The AAVSO Variable Star Index VSX (14) gives its spectral type as M5e-M9e.

X Oph was featured as Variable Star of the Year in the 2014 BAA Handbook (15). The article, written by John Toone, included a visual light curve of X Oph (Figure 1) with amplitude around 2 magnitudes and flat-bottomed minima at magnitude 8.4. These observations by Toone show the consistency expected of a good visual observer. The light of the Mira variable dominates when it is brighter than its constant companion but as it fades to minimum the light of the constant star dominates.

Based on a series of separation and position angle measurements of the components of X Oph between 1900 and 1957, Fernie (9) fitted an elliptical orbit and calculated a binary orbital period of 557 years. Including measurements up to 1975, Baize (16) calculated a period of 485 years. Further measurements led Novakovic (17) to give a period of 877 years in 2007. Given the relatively small fraction of the long highly eccentric orbit of X Oph observed so far, these differences are not surprising.

Clearly there is still uncertainty about many aspects of X Oph, but in particular about the spectral type of both its components and the magnitude of the constant star. X Oph was observed photometrically and spectroscopically by the author between 2016 May and December to see if a combination of these techniques could provide consistent values for the parameters of its component stars. In what follows the presumed Mira variable is referred to as the M star and the presumed constant star as the K star.

**Photometric observations**

Short series of filtered B and V magnitude measurements were made on 22 nights between 2016 May 24 and December 14 using a 0.35m SCT and SXVR-H9 CCD camera. Using B and V magnitudes of five nearby comparison stars from the APASS catalogue (18), ensemble photometry gave B and V magnitudes of X Oph which were transformed to the Johnson UBV standard system and are listed in Table 1 along with the corresponding B-V colour index. These measurements covered a period of time during which X Oph passed through a maximum around JD 2457548 (2016 June 8) and a minimum around JD 2457695 (2016 November 2).

Figure 2 shows magnitudes of X Oph submitted to the BAA Variable Star Section database (19) between 2015 November and 2017 January by Toone and the author. Visual estimates by Toone are shown in black and B and V measurements by the author in blue and green. During the descending part of the light curve the visual and V magnitudes agreed well while the visual estimates underestimated the peak V magnitude and overestimated the minimum V magnitude. Toone, in a private communication to the author, attributes this discrepancy to the change in apparent redness of the star between maximum and minimum. The V magnitude of the unresolved binary varied between 6.37 on 2016 May 29 and 8.59 on 2016 November 25. Variation of the B-V colour index over this period between 1.49 and 1.15 is shown in Figure 3. This change of colour arises from the changing relative brightness of two components with different spectral types as the Mira component fades and passes through minimum.

Given the measured magnitude of the binary (B), and an assumed magnitude of the K star (K), the corresponding magnitude of the M star (M) can be calculated using the formula

$$M = -2.5*\text{Log}_{10}(10^{(-0.4*B)} - 10^{(-0.4*K)}) \qquad (1)$$

Figure 4 shows measured V magnitudes for the binary and derived V magnitudes for the M star for a range of assumed V magnitudes for the K star. This indicates that practical bounds on the possible V magnitude of the constant K star are between 8.6, the brightest it can be given the observed binary minimum, and 9.2, which would result in an M star amplitude of 2.5 magnitudes, considered to be the smallest range for a Mira star.

**Spectroscopic observations**

Spectra of X Oph covering the wavelength range 3900 to 7400Å with spectral resolution 5 Å were recorded on 14 nights using a LISA spectrograph and SXVR-H694 CCD camera. The dates of these spectra are marked in Table 1 and on Figure 2. These spectra give the relative flux (spectral energy) across the spectrum.

In preparation for converting these to absolute flux spectra, absolute flux spectra of seventeen spectrophotometric standard stars from the CALSPEC HST Spectral Calibration Database (20) were convolved with the profile of the V filter to find the absolute flux transmitted by this filter and hence its mean spectroscopic zero point. This enabled conversion of a V magnitude to an absolute flux. Each measured relative flux spectrum was convolved with the spectral profile of the V filter to give the relative flux transmitted by the V filter. Knowing the V magnitude of the star on that date, each relative flux spectrum could then be scaled to produce an absolute flux spectrum in ergs/cm$^2$/s/Å.

The absolute flux spectrum of X Oph just before maximum recorded on 2016 May 29 is shown in Figure 5. This shows the characteristic saw-tooth pattern of TiO molecular absorption bands expected in the spectrum of a Mira variable. Also visible are Hγ and Hδ emission lines as reported by Merrill (6) around maximum light. By mid-August as the star faded these emission lines had disappeared.

**Determining component magnitudes and spectral types**

Adding spectra of the two components in absolute flux units gives the absolute flux spectrum of the binary. This offers a way to investigate the nature of the components of X Oph. Assuming a V magnitude and spectral type for each component, their individual absolute flux spectra can be synthesised using model spectra from the Pickles Stellar Spectral Flux Library (21) and combined to give a synthesised spectrum of the binary.

Figure 6 shows synthesised spectra for the K and M stars on 2016 November 10 around minimum light with their assumed parameter values and the combined binary spectrum. At these relative magnitudes the K star contributes more flux to the binary spectrum than the M star except at the red end of the spectrum.

The optimum choice of V magnitude and spectral type for each component on a specific date can be found by synthesising the absolute flux spectrum of the binary as described above and comparing it with the measured absolute flux spectrum of X Oph on that date. The best parameter values are those which give the smallest root mean square (rms) difference in flux between the synthesised and measured spectra over the wavelength range 3900Å to 7300Å. The effect of reddening on the observed spectrum is not clear but is probably relatively small given the value found by Fernie (9), and so has not been included in the analysis.

Previous analyses have indicated that, as a Mira variable, the likely spectral type of the M star is in the range M5III to M8III and that this may vary during the pulsation cycle. This variation of spectral type has been observed in other pulsating stars (22). Similarly, the K star spectral type is likely to be in the range K0III to K2III and its V magnitude in the range 8.6 to 9.2. Given the K magnitude, the magnitude of the M star can be found on any given date from equation (1). There are therefore three independent parameters.

Six dates with the highest quality spectra spanning the binary magnitude range from maximum to minimum were chosen, the three parameters varied within the above ranges, spectra synthesised for each component and the combined synthesised binary spectra compared with the measured spectra of X Oph on each date. Table 2 shows the parameter values which resulted in the best fits on each date. Figure 7 shows the best fitting synthesised and measured spectra for X Oph on 2016 May 29, August 23 and November 10. Note that there are different flux scales in each plot. Results for the other dates are similar.

Changing the assumed K star magnitude by 0.1 or the assumed spectral types of both components by one sub-unit (e.g M7 to M8) produced worse rms fits in each case. The largest mismatch between synthesised and measured spectra tends to be at the red end of the spectrum. This is where the change in the M star spectrum between sub-types is largest and as spectra are only being assigned to the nearest sub-type some degree of mismatch in this region is expected. In all cases changing the sub-type of the synthesised M star spectra produced a larger discrepancy than the ones shown.

**Conclusion**

Given the consistency of the results in Table 2, it seems likely that the V magnitude of the K star is 9.0 and its spectral type K1III. The spectral type of the M star changes from M6III at maximum to M7III as it fades and passes through minimum. As noted earlier, this variation is expected. With 9.0 as the V magnitude of the K star, the V magnitude of the M star varied from 6.47 at maximum to 9.83 at minimum, a range of 3.36 magnitudes and consistent with its identification as a Mira. Figure 8 shows V magnitude light curves of the X Oph binary and of the two components based on the parameter values in Table 2.

These results are consistent with previously published analyses and the use of flux calibrated spectra has provided a possible resolution of the earlier uncertainty about the properties of the components of X Oph. However a word of caution is in order as these results are based on analysing half of one pulsation cycle of the Mira variable and the long-term light curve of X Oph tells us that every cycle is subtly different.

**Acknowledgements**

I am grateful to John Toone for originally arousing my interest in this star and to both Toone and Chris Lloyd for providing references to useful background information. My thanks are due also to the referees whose constructive comments have helped to improve the paper.

Table 1. Standard Johnson B and V magnitudes and B-V colour index of X Oph on 22 nights between 2016 May 24 and December 14. Dates on which spectra were recorded are marked.

| Date | JD | B mag | B err | V mag | V err | B-V | B-V err | Spectra? |
|---|---|---|---|---|---|---|---|---|
| 2016 May 24 | 2457533.45802 | 7.906 | 0.058 | 6.414 | 0.074 | 1.492 | 0.109 | |
| 2016 May 29 | 2457538.48586 | 7.836 | 0.038 | 6.372 | 0.036 | 1.464 | 0.064 | Y |
| 2016 Jul 03 | 2457573.43727 | 8.371 | 0.021 | 6.953 | 0.023 | 1.418 | 0.033 | |
| 2016 Jul 05 | 2457575.48740 | 8.431 | 0.024 | 7.024 | 0.046 | 1.407 | 0.054 | Y |
| 2016 Jul 18 | 2457588.43082 | 8.721 | 0.019 | 7.356 | 0.021 | 1.365 | 0.035 | Y |
| 2016 Jul 24 | 2457594.42664 | 8.828 | 0.027 | 7.475 | 0.027 | 1.353 | 0.047 | |
| 2016 Aug 06 | 2457607.38447 | 9.105 | 0.021 | 7.753 | 0.027 | 1.352 | 0.039 | Y |
| 2016 Aug 08 | 2457609.38860 | 9.132 | 0.021 | 7.754 | 0.033 | 1.377 | 0.050 | Y |
| 2016 Aug 22 | 2457623.47365 | 9.359 | 0.019 | 8.039 | 0.020 | 1.320 | 0.033 | |
| 2016 Aug 23 | 2457624.35873 | 9.399 | 0.020 | 8.085 | 0.008 | 1.314 | 0.023 | Y |
| 2016 Sep 11 | 2457643.31395 | 9.509 | 0.022 | 8.282 | 0.013 | 1.226 | 0.027 | Y |
| 2016 Sep 13 | 2457645.37313 | 9.482 | 0.024 | 8.314 | 0.016 | 1.168 | 0.031 | |
| 2016 Sep 16 | 2457648.43751 | 9.558 | 0.013 | 8.375 | 0.024 | 1.184 | 0.023 | Y |
| 2016 Sep 22 | 2457654.31841 | 9.526 | 0.014 | 8.371 | 0.013 | 1.155 | 0.021 | Y |
| 2016 Sep 25 | 2457657.37999 | 9.571 | 0.025 | 8.388 | 0.020 | 1.183 | 0.038 | Y |
| 2016 Oct 22 | 2457684.34154 | 9.680 | 0.021 | 8.524 | 0.025 | 1.156 | 0.035 | |
| 2016 Oct 31 | 2457693.30471 | 9.714 | 0.030 | 8.549 | 0.034 | 1.165 | 0.034 | Y |
| 2016 Nov 10 | 2457703.23880 | 9.733 | 0.007 | 8.582 | 0.011 | 1.151 | 0.014 | Y |
| 2016 Nov 25 | 2457718.22946 | 9.780 | 0.013 | 8.585 | 0.019 | 1.195 | 0.020 | Y |
| 2016 Nov 30 | 2457723.25896 | 9.711 | 0.036 | 8.485 | 0.032 | 1.226 | 0.048 | |
| 2016 Dec 01 | 2457724.21967 | 9.691 | 0.013 | 8.502 | 0.016 | 1.188 | 0.019 | Y |
| 2016 Dec 14 | 2457737.23001 | 9.601 | 0.030 | 8.365 | 0.050 | 1.236 | 0.056 | |

Table 2. Measured V magnitudes of X Oph and the parameter values of its components which give the best fit between synthesised and measured absolute flux spectra on the dates shown.

| Date | X Oph V magnitude (measured) | K star V magnitude (fitted) | M star V magnitude (calculated) | K star spectral type (fitted) | M star spectral type (fitted) |
|---|---|---|---|---|---|
| 2016 May 29 | 6.37 | 9.0 | 6.47 | K1 | M6 |
| 2016 Jul 18 | 7.36 | 9.0 | 7.63 | K1 | M7 |
| 2016 Aug 23 | 8.09 | 9.0 | 8.70 | K1 | M7 |
| 2016 Sep 22 | 8.39 | 9.0 | 9.30 | K1 | M7 |
| 2016 Nov 10 | 8.58 | 9.0 | 9.82 | K1 | M7 |
| 2016 Dec 01 | 8.50 | 9.0 | 9.59 | K1 | M7 |

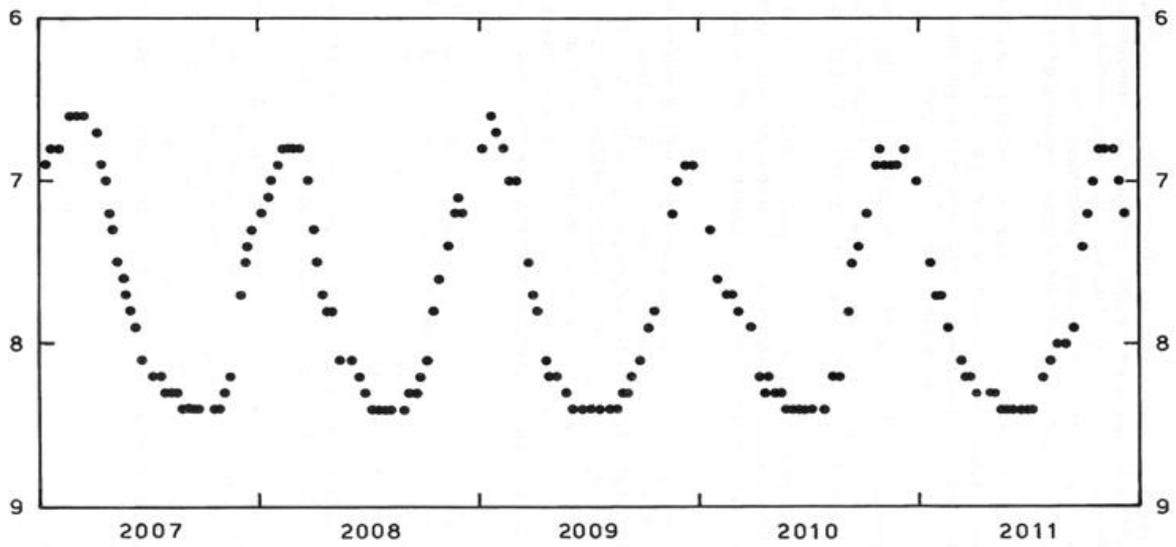

Figure 1. Visual light curve of X Oph from the 2014 BAA Handbook.

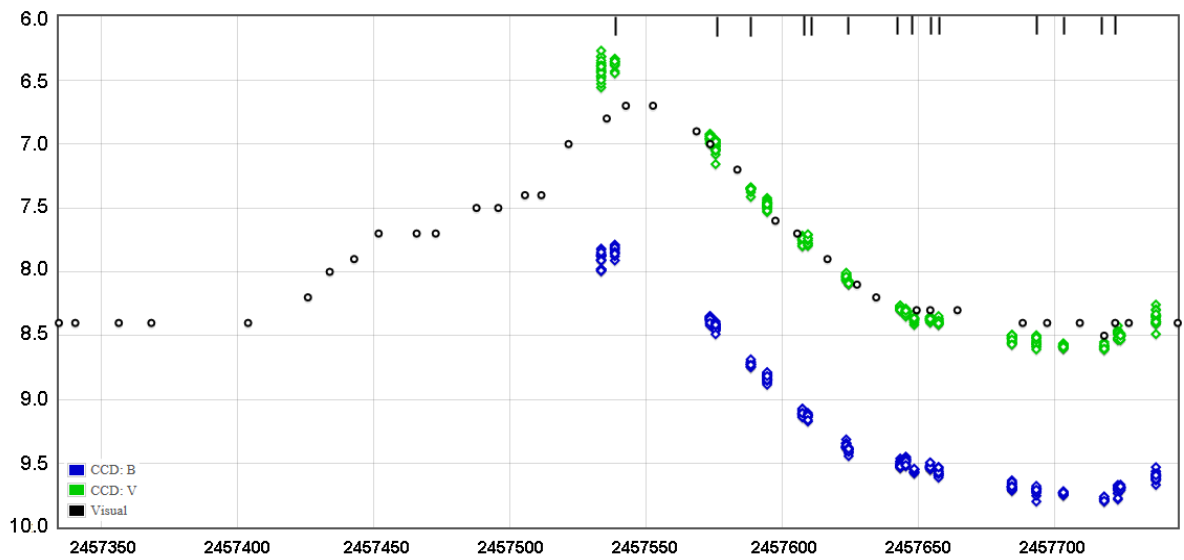

Figure 2. Visual, B and V magnitudes of X Oph in the BAA VSS database contributed by Toone (black) and the author (blue and green) between 2015 November and 2017 January. The dates on which spectra were recorded are marked at the top.

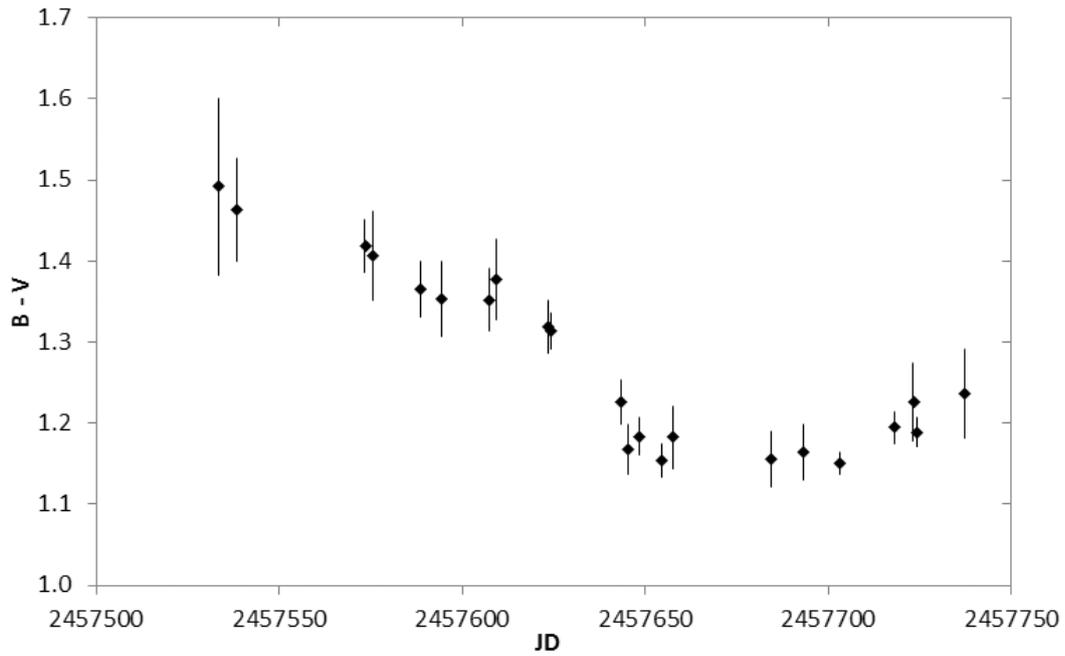

Figure 3. B-V colour index of X Oph between 2016 May and December.

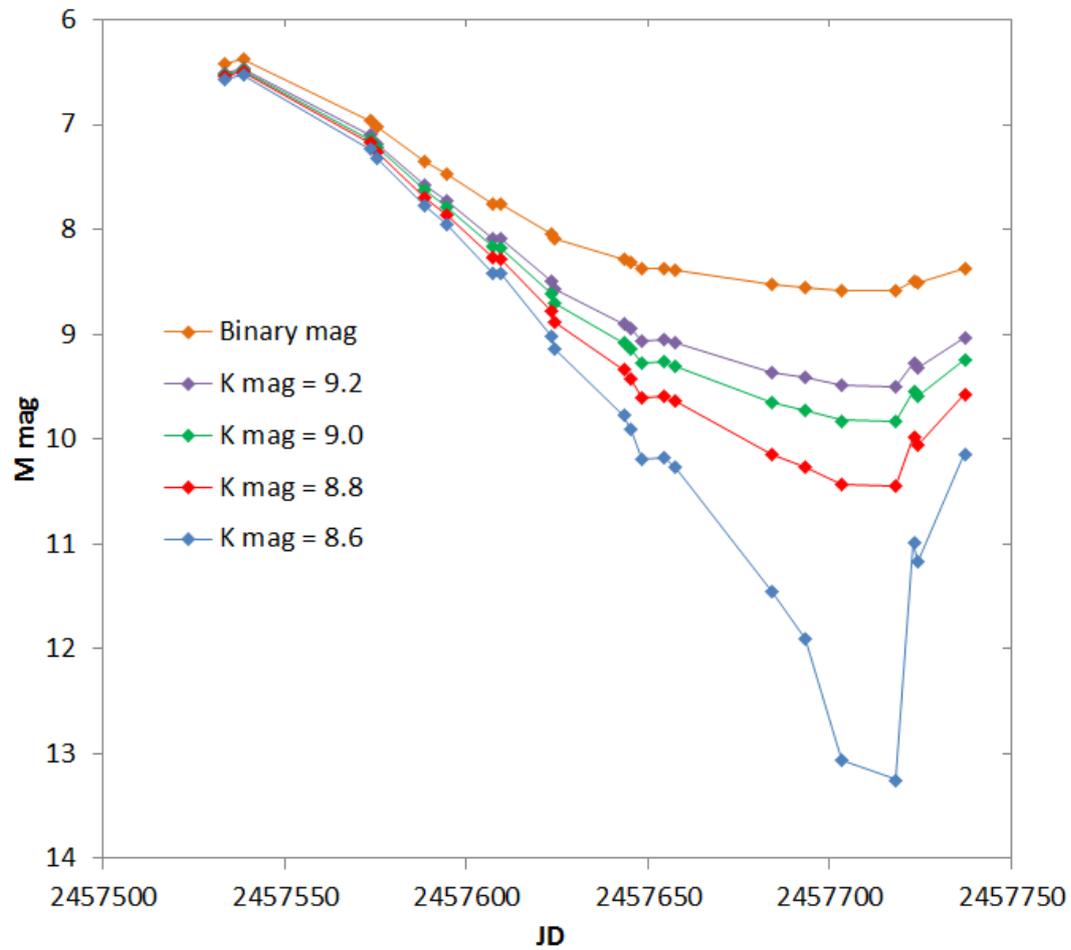

Figure 4. Derived magnitudes for the M star for a range of assumed magnitudes for the K star given the measured binary magnitudes.

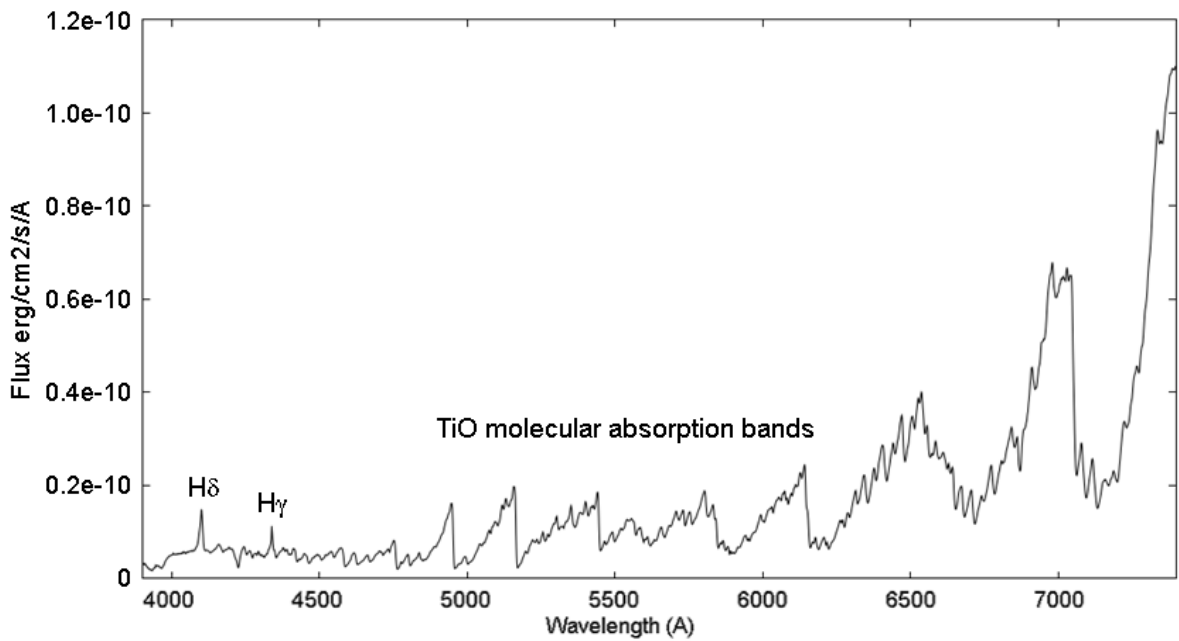

Figure 5. Absolute flux spectrum of X Oph recorded on 2016 May 29 around maximum light. TiO molecular absorption bands typical of a Mira variable are prominent as are Hγ and Hδ emission lines.

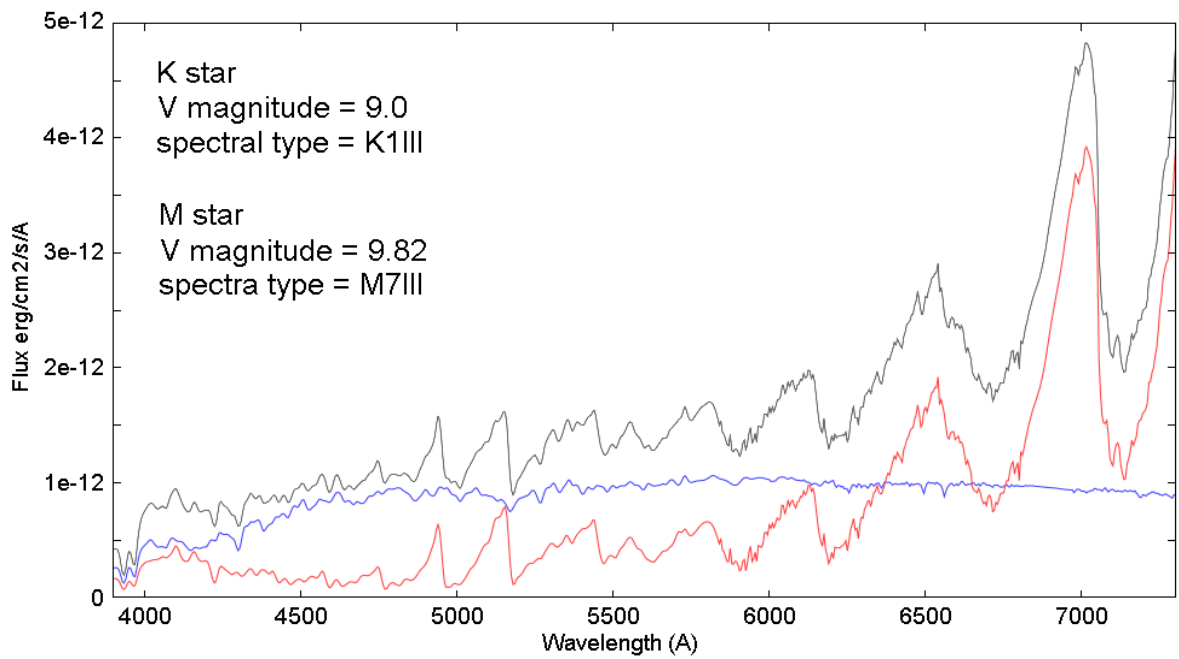

Figure 6. Synthesised spectra for the K star (blue) and M star (red) on 2016 November 10 around minimum light with their assumed parameter values and the combined binary spectrum (black).

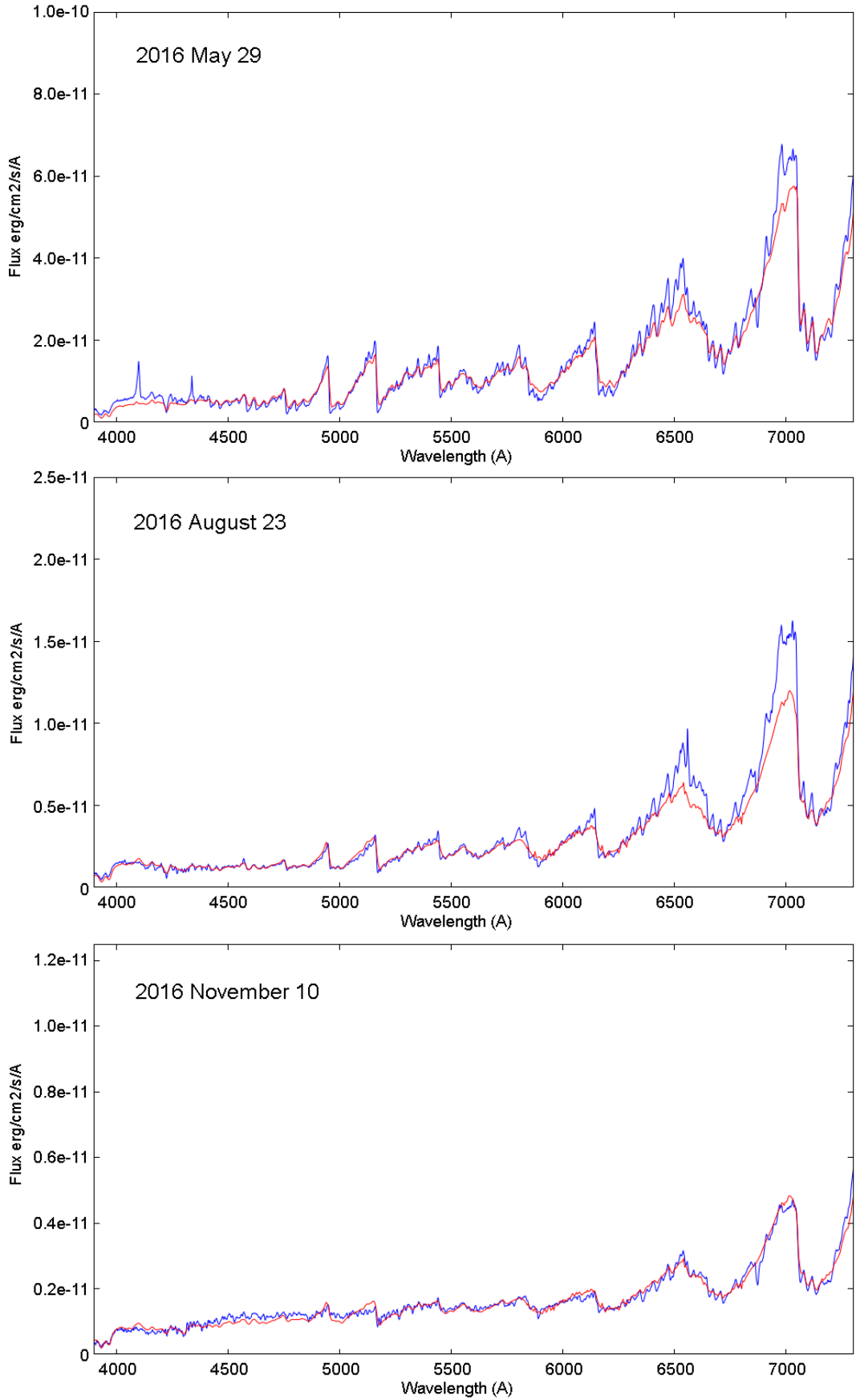

Figure 7. Best fitting synthesised (red) and measured (blue) spectra for X Oph on 2016 May 29, August 23 and November 10. Note flux scales are different in each plot.

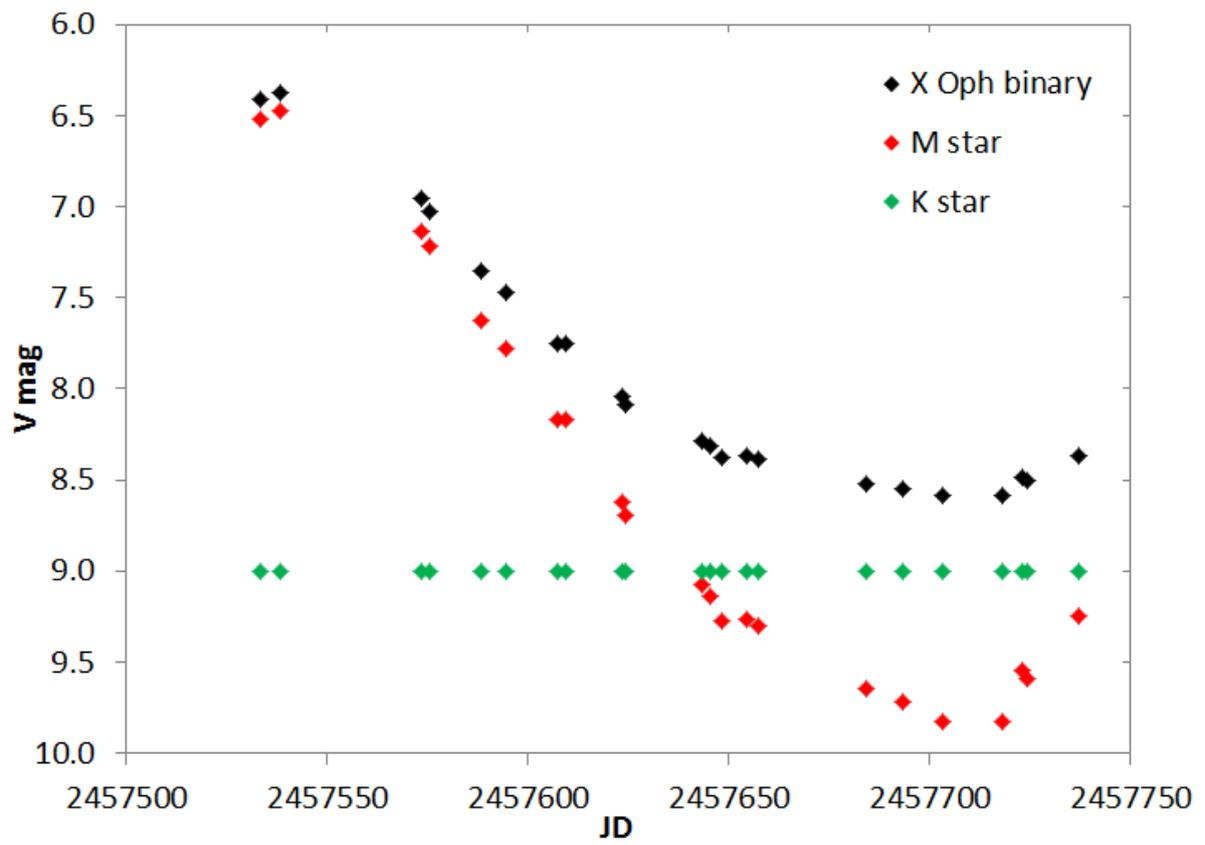

Figure 8. V magnitude light curves of the X Oph binary system and its components based on the parameter values in Table 2.